\documentclass[acmsmall]{acmart}

\setcopyright{cc}
\setcctype{by}
\acmDOI{10.1145/3715760}
\acmYear{2025}
\acmJournal{PACMSE}
\acmVolume{2}
\acmNumber{FSE}
\acmArticle{FSE043}
\acmMonth{7}
\received{2024-09-13}
\received[accepted]{2025-01-14}

\usepackage{pifont}
\usepackage{cleveref}
\usepackage{multirow}
\usepackage{subcaption}
\usepackage{caption}
\DeclareCaptionType{Prompt}
\usepackage{enumitem}
\setlist{leftmargin=3.7mm}
\usepackage{xcolor}
\usepackage{listings}
\definecolor{halfgray}{gray}{0.35}
\definecolor{deepblue}{rgb}{0,0,0.5}
\definecolor{deepred}{rgb}{0.6,0,0}
\definecolor{deepgreen}{rgb}{0,0.5,0}
\definecolor{highlightorange}{rgb}{0.98, 0.92, 0.8}

\makeatletter
\newcommand\HL{%
	\gdef\lst@alloverstyle##1{%
		\fboxrule=0pt
		\fboxsep=0pt
		\colorbox{lightgray}{\strut##1}%
	}%
}
\newcommand\HLoff{%
	\xdef\lst@alloverstyle##1{##1}%
}

\definecolor{lightgreen}{rgb}{0.9,1,0.9}
\definecolor{lightred}{rgb}{1,0.9,0.9}

\makeatletter
\newcommand\HLgreen{%
	\gdef\lst@alloverstyle##1{%
		\fboxrule=0pt
		\fboxsep=0pt
		\colorbox{lightgreen}{\strut##1}%
	}%
}
\newcommand\HLgreenoff{%
	\xdef\lst@alloverstyle##1{##1}%
}

\makeatletter
\newcommand\HLred{%
	\gdef\lst@alloverstyle##1{%
		\fboxrule=0pt
		\fboxsep=0pt
		\colorbox{lightred}{\strut##1}%
	}%
}
\newcommand\HLredoff{%
	\xdef\lst@alloverstyle##1{##1}%
}

\newcommand{\code}[1]{{\ttfamily \small #1}}

\newcommand{\name}{ChangeGuard}

\begin{document}

\title{\name{}: Validating Code Changes via Pairwise Learning-Guided Execution}

\author{Lars Gröninger}
\affiliation{
  \institution{University of Stuttgart}
  \country{Germany}
}
\email{lars.groninger@gmail.com}
\author{Beatriz Souza}
\affiliation{
  \institution{University of Stuttgart}
  \country{Germany}
}
\email{beatrizbzsouza@gmail.com}
\author{Michael Pradel}
\affiliation{
  \institution{University of Stuttgart}
  \country{Germany}
}
\email{michael@binaervarianz.de}

\begin{abstract}
    Code changes are an integral part of the software development process.
    Many code changes are meant to improve the code without changing its functional behavior, e.g., refactorings and performance improvements.
    Unfortunately, validating whether a code change preserves the behavior is non-trivial, particularly when the code change is performed deep inside a complex project.
    This paper presents \name{}, an approach that uses learning-guided execution to compare the runtime behavior of a modified function.
    The approach is enabled by the novel concept of pairwise learning-guided execution and by a set of techniques that improve the robustness and coverage of the state-of-the-art learning-guided execution technique.
    Our evaluation applies \name{} to a dataset of 224 manually annotated code changes from popular Python open-source projects and to three datasets of code changes obtained by applying automated code transformations.
    Our results show that the approach identifies semantics-changing code changes with a precision of 77.1\% and a recall of 69.5\%, and that it detects unexpected behavioral changes introduced by automatic code refactoring tools.
    In contrast, the existing regression tests of the analyzed projects miss the vast majority of semantics-changing code changes, with a recall of only 7.6\%.
    We envision our approach being useful for detecting unintended behavioral changes early in the development process and for improving the quality of automated code transformations.
\end{abstract}

\ccsdesc[500]{Software and its engineering~Software verification and validation}
\ccsdesc[500]{Software and its engineering~Software defect analysis}
\ccsdesc[300]{Software and its engineering~Software evolution}
\ccsdesc[300]{Software and its engineering~Software maintenance tools}

\keywords{Code changes, refactoring, differential testing}

\maketitle

\section{Introduction}

Successful software projects are continuously evolving.
Developers implement new features, fix existing bugs, refactor code to increase its readability, or optimize the performance of a frequently executed code piece.
Such changes are performed either by human developers or by automated tools.
Regardless of who or what performs a code change, all changes are made to achieve a specific goal, e.g., to fix a bug, improve performance, or make the changed code more readable.
Depending on the goal, the semantics of the affected code is supposed to change in a particular way or to not change at all.
For example, a refactoring or a performance improvement should not change the semantics, while a bug fix should change the behavior so that the new behavior matches the desired, correct behavior.

Determining whether a given code change alters the semantics of the code is undecidable in theory and far from trivial in practice.
One option is for a human to statically inspect the code change and reason about its impact on the behavior of the affected code.
However, this approach is time-consuming, error-prone, and does not scale.
Another option is to rely on a regression test suite to exercise the code before and after the change.
However, this approach is only feasible if there exist test cases that exercise the changed code locations with various inputs and provide assertions able to detect any behavioral differences.
Overall, because validating code changes is difficult, developers may accidentally introduce bugs by making changes that unintentionally modify the behavior of the affected code.

\begin{figure}
    \centering
    \begin{lstlisting}
        def _retry(self, request, reason, spider):
            retries = request.meta.get('retry_times', 0) + 1
        (@\HLred@)-   retry_times = self.max_retry_times(@\HLredoff@)
        (@\HLred@)-   if 'max_retry_times' in request.meta: retry_times = request.meta['max_retry_times'](@\HLredoff@)
        (@\HLgreen@)+   retry_times = request.meta.get('max_retry_times') or self.max_retry_times(@\HLgreenoff@)
            stats = spider.crawler.stats
            if retries <= retry_times:
                # (19 more, unchanged lines)
    \end{lstlisting}
    \caption{Motivating example of a code change meant to be semantics-preserving.}
    \label{fig:motivating}
\end{figure}
  
  As a motivating example, consider the code change in \Cref{fig:motivating}.
  As evidenced by the commit message ``Simplify retry\_times assignment statement'', the developers intended to improve the code without changing its behavior.\footnote{\url{https://github.com/scrapy/scrapy/commit/694c6d3d}}
  However, careful reasoning reveals that the code change actually modifies the behavior of the affected code:
  If the \code{'max\_retry\_times'} of \code{request.meta} is \code{0}, then \code{retry\_times} is set to \code{0} in the old version, but to \code{self.max.retry\_times} in the new version.
  This change is clearly unintended, yet no existing technique detected it.
  Interestingly, the developers realized their mistakes several days after the first code change and fixed it.\footnote{\url{https://github.com/scrapy/scrapy/commit/49c5afc5}}
  To detect such mistakes earlier, it would be beneficial to have an automated technique that can reason about the behavior of code changes and identify those that modify the behavior of the changed code.

An approach for automatically comparing the behavior of two versions of a code snippet must address two key challenges.
The first challenge is to actually execute the changed code, ideally with a diverse set of inputs, to cover as many execution paths as possible.
Since code changes may affect arbitrary code locations deep inside a complex project, reaching the changed code from the project's entry point(s) is non-trivial.
The second challenge is comparing the behavior of the two versions of the code snippet.
This comparison must be able to detect even subtle differences in the behavior, such as changes in the return value, the output printed to the console, the functions called, or the exceptions raised.
To the best of our knowledge, there currently exists no approach that can automatically compare two versions of a code snippet and determine whether the code change is semantics-preserving or semantics-changing.

This paper presents \name{}, an approach that compares two versions of a code snippet to determine whether the code change is semantics-preserving or semantics-changing.
The approach builds on the recently introduced concept of learning-guided execution~\cite{fse2023-LExecutor,icse2025-Treefix}, which allows executing arbitrary code snippets in isolation by predicting and injecting otherwise missing values.
We introduce the new concept of \emph{pairwise learning-guided execution}, which executes two versions of a code snippet side-by-side, while predicting and injecting any missing values to enable executing the code snippets in isolation.
The approach is enabled by a set of techniques to inject diverse, project-specific, and realistic values, to ensure consistency and non-interference between the two executions, and to handle calls to external functions and indexing operations.
For the motivating example in \Cref{fig:motivating}, the approach starts to execute both versions of the code at the beginning of the surrounding function, injects several values that would usually be missing and cause the code to crash, until reaching the changed code lines.
By comparing the behavior of the modified function before and after the change, \name{} determines that the change modifies the observable behavior of the function, which could have helped the developers to identify this issue earlier.

We evaluate \name{} on four newly gathered datasets of code changes in Python code: a set of 224 code changes extracted from ten open-source projects, which we manually label as semantics-preserving or semantics-changing, and three sets of code changes obtained by applying automated code transformations proposed by a rule-based and two LLM-based tools.
Our results show that \name{} effectively identifies semantics-changing code changes with a precision of 77.1\% and a recall of 69.5\%.
In contrast, the existing regression tests of the projects provide only 7.6\% recall, i.e., they miss the majority of semantics-changing code changes.
We also apply \name{} to code changes created via automated, supposedly semantics-preserving code transformations, many of which the approach shows to be semantics-changing.
We find that this effectiveness critically depends on our improvements over the state-of-the-art learning-based execution technique~\cite{fse2023-LExecutor}, which increase the median coverage of the changed code from 27\% to 92\%.
Finally, we evaluate the efficiency of \name{} and show that it finds semantics-changing behavior within a few seconds.

\name{} is the first to adapt and systematically evaluate learning-guided execution on pairs of code snippets and the practically relevant task of reasoning about code changes.
Prior work on learning-guided execution~\cite{fse2023-LExecutor} provides a preliminary evaluation on code changes.
However, their evaluation on code changes does not compare against a ground truth.
Our evaluation shows that out-of-the-box learning-guided execution fails to correctly classify many code changes, e.g., because it does not cover the code paths affected by the change.
Our work also relates to efforts toward automated reasoning about code changes, such as heuristics to guide symbolic execution toward modified code~\cite{Marinescu2013}.
\name{} differs by building on learning-guided execution, which creates more realistic input values and is effective at covering the vast majority of changed code.
Another related stream of work is on just-in-time defect prediction~\cite{Kamei2013,Hoang2019,DBLP:journals/csur/ZhaoDC23}.
Unlike our work, it aims at determining whether a code change is likely to introduce a defect, rather than whether the code change is semantics-preserving.
Finally, this paper also relates to work on identifying equivalent code, e.g., by mining functionally equal code fragments via random testing~\cite{Jiang2009a} or by checking whether two functions in binaries are semantically similar~\cite{Egele2014,binary2}.
In contrast to those efforts, \name{} focuses on semantic changes introduced by code changes.

In summary, our contributions are as follows:
\vspace{-1.4mm}
\begin{itemize}
  \item Pairwise learning-guided execution, a novel technique for comparing two code snippets.
  \item Techniques that significantly extend the robustness and coverage of learning-guided execution.
  \item A novel dataset of 224 manually annotated code changes from ten popular open-source projects.
  \item Empirical evidence that \name{} can accurately identify whether a given code change is semantics-preserving or semantics-changing.
\end{itemize}

\section{Approach}

The following presents our \name{} approach.
Given a code change that modifies a function $f$ from $f_{old}$ to $f_{new}$, the goal is to determine whether the change preserves the input-output semantics of the function.
We consider the code change to be semantics-changing if there exists a set of input values used by $f$ that causes $f_{old}$ to produce different observable behavior than $f_{new}$.

\subsection{Overview}

\begin{figure}
  \centering
  \includegraphics[width=0.8\columnwidth]{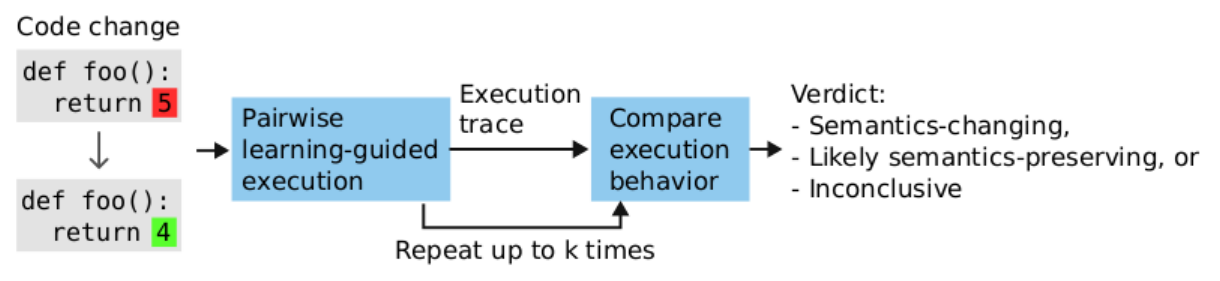}
  \caption{Overview of \name{}.}
  \label{fig:overview}
  \vspace{-1.5mm}
\end{figure}

Figure~\ref{fig:overview} shows an overview of our approach.
Given a code change, the approach determines how the change affects the semantics of the function.
The approach consists of two phases.
The first phase performs pairwise learning-guided execution, which is the main technical contribution of this work.
In a nutshell, pairwise learning-guided execution executes the old and the new version of the function, while injecting any missing values to enable executing the functions in isolation.
The second phase compares the execution behavior of the two versions of the function and reports the code change to be either semantics-changing, likely semantics-preserving, or in case the approach fails to perform a meaningful comparison, indicates an inconclusive outcome.
Because a single pairwise learning-guided execution reasons about only one of potentially many possible executions, the approach repeats the two phases until it determines the code change to be semantics-changing or until exceeding a configurable limit of $k$ executions.

\subsection{Background: Learning-Guided Execution}
\label{sec:background}

Before going into the details of our approach, we provide the necessary background on learning-guided execution~\cite{fse2023-LExecutor}, on which \name{} builds.
The key idea of learning-guided execution is to enable the execution of otherwise unexecutable code by predicting any values that are not initialized in the code by querying a machine learning model during the code execution.
The machine learning model is fine-tuned version of the pre-trained CodeT5 model~\cite{codet5}.
To support learning-guided execution, the model is fine-tuned on a dataset of ``normal'' executions on the following task: Given a piece of code that accesses a variable, attribute, or calls a function, predict the value that the code would read or observe as the return value of the function call.
Learning-guided execution is implemented via source-to-source instrumentation of the code to be executed.
The instrumentation intercept all reads of variables, reads of attributes, and calls to functions.
If the given code accesses an initialized value or calls an existing function, the instrumented code continues with the regular execution.
If, instead, the code accesses an uninitialized value or calls a non-existing function, i.e., the code would usually crash, the instrumented code queries the machine learning model and injects the returned value into the running program.
By injecting a (return) value that is likely realistic for the given code, the execution can continue and the code can be executed in isolation.

Prior work has proposed learning-guided execution as a general concept to execute individual code snippets~\cite{fse2023-LExecutor}.
Our work is the first to adapt and systematically apply this concept to pairs of code snippets and the practically relevant task of reasoning about code changes.

\subsection{Pairwise Learning-Guided Execution}
\label{sec:pairwise lexecution}

The following presents the core technical contribution of this work: pairwise learning-guided execution.
The basic idea is to execute two versions of a code snippet side-by-side, while predicting and injecting any missing values into the execution.
Like learning-guided execution, the approach is designed to execute possibly incomplete code snippets, i.e., code snippets that would likely crash due to missing values when being executed as-is.
Unlike existing work on learning-guided execution, we here target a pair of code snippets, instead of a single code snippet.
Specifically, we focus on the task of comparing two versions of a function, $f_{old}$ and $f_{new}$, to determine whether the code change is semantics-preserving or semantics-changing.
The following describes the challenges that arise during pairwise learning-guided execution and how \name{} addresses them.

\subsubsection{Merging Functions into a Comparison Program}

To prepare the given function versions for pairwise learning-guided execution, we first merge them into a single program.
The motivation for this step is to enable the execution of the two functions on the same values and to enable the approach to directly compare their return values.
Given two functions $f_{old}$ and $f_{new}$, we construct a program that consists of three parts.
First, the program contains the two function definitions, where each function is given a unique name, e.g., \code{f\_old} and \code{f\_new}.
The approach removes the formal parameters of the functions, so the functions can be called without arguments.
Any usage of a formal parameter in the function body will be handled by injecting values during the execution, following the concept of learning-guided execution.
Second, the program contains a call to the old function, \code{f\_old()}, followed by a call to the new function, \code{f\_new()}.
Finally, the program contains code to compare the execution behavior of the two functions, which we describe in Section~\ref{sec:comparison}.

\subsubsection{Injecting Diverse, Project-specific, and Realistic Values}
\label{sec:injecting values}

Learning-guided execution is based on predicting and then injecting otherwise missing values.
The nature of these values is crucial for the success of the execution.
However, as we show in our evaluation, using the existing learning-guided execution technique~\cite{fse2023-LExecutor} out-of-the-box fails to cover many of the changed code lines, and hence, fails to provide a meaningful comparison of many function pairs.
A crucial reason for this limitation is that the values injected by the existing learning-guided execution approach are sampled from only 23 concrete values, such as \code{1}, \code{"a"}, \code{True}, \code{False}, \code{None}, an empty list, or a list consisting of a single, simple object.
While these values are sufficient to execute some code snippets without crashing, they often fail to cover branches guarded by non-trivial conditions or cause the function to crash before reaching the changed code locations.

\begin{table}
    \caption{Abstract values and their concretizations.}
    \label{tab:values}
    \begin{tabular}{@{}lp{32em}@{}}
      \toprule
      Abstract value & Concrete value(s)                                                                                                               \\
      \midrule
      None           & \code{None}                                                                                                                     \\
      Boolean        & \code{True}, \code{False}                                                                                                       \\
      Integer        & \code{-100}, \code{-10}, \code{-1}, \code{0}, \code{1}, \code{10}, \code{100}, and all integer literals in the code             \\
      Float          & \code{-100.0}, \code{-10.0}, \code{-1.0}, \code{0.0}, \code{1.0}, \code{10.0}, \code{100.0}, and all float literals in the code \\
      String         & \code{""}, \code{"a"}, and all string literals in the code                                                                      \\
      List           & List with random size and elements of a randomly selected type                                                       \\
      Tuple          & Tuple with random size and elements of a randomly selected type                                                      \\
      Dictionary     & Dictionary with random size that maps strings to values of a random type                                  \\
      Set            & Set with random size and elements of a randomly selected type                                                        \\
      Callable       & Class of a versatile object with various default methods                                                                                \\
      Resource       & Versatile object with various default methods                                                                                \\
      Object         & Versatile object with various default methods                                                                                \\
      \bottomrule
    \end{tabular}
\end{table}

\name{} addresses this limitation by extending the existing learning-guided execution approach with a set of techniques to inject diverse, project-specific, and realistic values, as presented in the following.
The basic idea is to let the neural model predict one of twelve abstract values to inject at a specific code location, and to then concretize the predicted abstract value to a concrete value sampled from a (theoretically infinitely) large pool of possible values.
Table~\ref{tab:values} shows the twelve abstract values and their concretizations.
The abstract values correspond to the ``coarse-grained abstraction'' in prior work~\cite{fse2023-LExecutor}.
Unlike the prior work, which concretizes the abstract values into one of only 23 fixed values, \name{} samples from many more values.
This pool contains a diverse set of hard-coded primitive values, as shown in the table.
More importantly, the pool contains project-specific literals, a versatile object with rich default behavior, and randomly constructed data structures, as presented in the following.

The architecture and training process of the neural model is the same as in prior work~\cite{fse2023-LExecutor}, i.e., a fine-tuned CodeT5~\cite{codet5} model (Section~\ref{sec:background}).
Section~\ref{sec:setup fine-tuning} provides more details on the training data.

\paragraph{Project-specific literals.}

To increase the diversity of values and enable our approach to inject values that are specific to the project under test, we extract literals from the code under test.
To this end, \name{} parses the old and the new function and extracts all integer, float, and string literals.
For each of these three types, the extracted literals are added to the pool of concrete values of that type.
When executing code and injecting a value of a specific type, \name{} randomly samples from all values in the pool of concrete values of that type.
For example, suppose a function contains a check \code{if val == "new"}, then the approach will add the string \code{"new"} to the pool of strings to concretize to, which enables \name{} to execute the branch guarded by this check.

\paragraph{Generating diverse data structures.}

A commonly missing kind of value are complex data structures, such as lists, sets, or dictionaries.
To vary the size and the content of these data structures, \name{} randomly generates complex values in the following way.
At first, the approach determines the size of the data structure by randomly sampling a value between zero and four.
The rationale for this heuristic choice is to support sequence unpacking\footnote{\url{https://docs.python.org/3/tutorial/datastructures.html\#tuples-and-sequences}} and the fact that unpacking sequences of sizes larger than four is uncommon in practice.
Next, \name{} randomly selects the type of values to add into the data structure, where objects are created with 50\% probability and otherwise, the approach randomly decides between integers, floats, strings, boolean, and \code{None}.
The reason for biasing the selection toward objects is that, as explained below, \name{} injects objects with a rich set of default behaviors.
Finally, the approach fills the data structure with concrete values created by recursively invoking the value concretization algorithm.

For example, consider a code snippet that reads a variable \code{my\_dict} for which the neural model predicts the abstract value ``dictionary''.
\name{} will create a dictionary with a randomly selected size and with randomly initialized key-value pairs, e.g., \code{\{"a": 42, "special": set()\}}.

\paragraph{Versatile objects with rich default behavior.}

Whenever the neural model predicts the abstract value to be a resource or an object, \name{} injects an instance of a special class designed to provide a default object with rich behavior, which we call the \emph{versatile object}.
The goal of this class is to mimic the behavior of a wide range of real-world objects, such as built-in types, user-defined classes, or objects from third-party libraries.
To this end, the versatile object implements most of Python's special methods\footnote{\url{https://docs.python.org/3/reference/datamodel.html\#specialnames}}, such as \code{\_\_add\_\_}, \code{\_\_getitem\_\_}, \code{\_\_xor\_\_}, and \code{\_\_enter\_\_}, with a default behavior designed to not stop the execution.
For example, let \code{o} be a variable predicted to be an object, which is then used in a statement \code{x = o + 3}.
The class of the versatile object implements the \code{\_\_add\_\_} method so that it determines that the statement is supposed to be an integer addition (as opposed to, e.g., a string concatenation) based on the value of the other operand, i.e., \code{3}, and then uses its internal integer representation, e.g., the value \code{1} to perform the addition.
As a result, the statement will assign \code{4} to \code{x}, and the program continues to execute.
Based on such implementations of the special methods, a versatile object can be used in a wide range of scenarios, such as addition statements, indexing operations, and when being used as a resource.

\subsubsection{Ensuring Consistency and Non-Interference}
One of the key challenges for pairwise learning-guided execution is to ensure that the executions of the two functions are consistent without interfering with each other.
Consistency here means that both functions should operate on the same values, which is essential to compare their behavior.
Non-interference here means that the execution of one function should not affect the execution of the other function, which is important to ensure that any observed behavioral difference is due to the code change and not due to the interaction between the two functions.
\name{} ensures these two properties by maintaining a consistency map and by handling renamed variables, as described in the following.

\paragraph{Consistency map with deep copies.}
To ensure both consistency and non-interference, \name{} maintains a \emph{consistency map} that maps each predicted variable or attribute to a pair of values.
The keys of the map are the access path used to refer to the variable or attribute, e.g., \code{data.ages}.
The access path uniquely identifies the variable or attribute, unlike prior work~\cite{fse2023-LExecutor}, which tries to ensure consistency using the unqualified name of attributes only.
Each key in the consistency maps to a pair $(v_{old}, v_{new})$ of values, where the second element $v_{new}$ is a deep copy of the first element $v_{old}$.
For example, suppose a function is reading an undefined attribute \code{data.ages} and \name{} injects a mutable list created via \code{[23, 42]}.
The consistency map would then contain the entry \code{data.ages} $\rightarrow$ (\code{[23, 42]}, \code{deepcopy([23, 42])}), where \code{deepcopy} is a function provided by the builtin \code{copy} module of Python.
The first value of the pair is injected into the execution of the $f_{old}$ function, whereas the second value is injected into the execution of the the $f_{new}$ function.
The reason for using a pair of values is to prevent modifications of a value in one function from affecting uses of the value in the other function.
Without this separation, the execution of $f_{old}$ could modify the value, e.g., by appending an element to a list, which would then affect the execution of $f_{new}$.

\paragraph{Handling renamed variables.}
The consistency map uses access paths, i.e., fully qualified identifier names, to keep the injected values consistent across both executions.
However, this approach falls short if an identifier has been renamed as part of the analyzed code change.
Naively handling a renamed variable as two separate variables would cause the approach to inject different values, which could lead to detecting diverging behavior even though nothing but the name has changed.
For example, consider a code change that renames the variable \code{data} to \code{store}.
Because the old and the new variable names are different, the approach would create and inject different values for the two variables, which may then cause the two functions to appear to behave differently.
To avoid such spurious changes in behavior, we enhance the consistency map to handle renamed variables.
Given a mapping of old names to new names, \name{} merges the old and the new variable name into a new unique name and then uses the merged name as the key in the consistency map.
For the above example, the consistency map would contain an entry \code{data\_renamed\_store} $\rightarrow$ ($v_{old}, v_{new}$).
When the old name is encountered during the execution of the old function $f_{old}$, the approach looks up the merged name in the map and returns the first element $v_{old}$, and likewise for the new function.
For our experiments, we manually create the mapping of old names to new names, but we envision this to be automated~\cite{DBLP:journals/tse/TsantalisKD22} for a real-world deployment.

\subsubsection{Handling Calls to External Functions}

Because \name{} executes the two functions in isolation, it needs to handle calls to other functions, e.g., functions imported in a third-party project.
Our approach follows the general idea of learning-guided execution, which is to predict and inject return values for calls to external functions.
We improve this idea in three ways.

\begin{figure}
  \begin{lstlisting}
  try:
      parse(code)
  except SyntaxError as e:
      print("Caught exception:", e)
  \end{lstlisting}
  \caption{Example of handling exceptions triggered by external calls.}
  \label{fig:exception handling}
\end{figure}

\paragraph{Inject exceptions thrown by external functions.}
The first change is motivated by the fact that calls to external functions may raise exceptions, and that the analyzed function may have code to handle these exceptions.
With the existing learning-guided execution approach~\cite{fse2023-LExecutor}, calls to external functions never raises an exception, and hence, code that handles such exceptions would never be executed.
Instead, \name{} tries to mimic exceptional behavior that external functions may trigger.
To this end, the approach statically extracts the types of exceptions that are caught in a try-except statement and associates them with function calls in the corresponding try block.
Whenever the approach reaches a call to an undefined function, then with a configurable probability (default: 15\%), the approach triggers the corresponding exception.
The exception is then caught by the surrounding try-except statement, allowing \name{} to reason about any differences in how the two functions handle exceptions.
For example, consider the code in \Cref{fig:exception handling}, where calling \code{parse} may raise a \code{SyntaxError}, and assume that the \code{parse} function is not defined within the analyzed function.
\name{} associates the \code{parse} function with the \code{SyntaxError} exception, and when the approach reaches the call of \code{parse}, it will probabilistically raise a \code{SyntaxError}.
As a result, the code in the except block will be executed, and the approach can reason about any differences in the output printed by the two functions.

\paragraph{Type checks.}
The second change is to replace calls to \code{isinstance} with a call to a custom function.
To illustrate the motivation, consider an expression \code{isinstance(x, ClassA) and isinstance(x, ClassB)} in the analyzed function.
Assuming \code{ClassA} and \code{ClassB} are not in a subtype relationship, the expression should always evaluate to false.
However, if the approach injects a versatile object for \code{x} and injects the class of the versatile object both for the \code{ClassA} and \code{ClassB}, then the expression would evaluate to true.
In other words, because the isolated code execution does not have access to the actual classes, the \code{isinstance} checks may not behave as expected.
To address this problem, \name{} statically extracts all classes that appear in any \code{isinstance} call in the analyzed functions and randomly assigns one of these types each time one of our versatile objects is created.
Our custom replacement function then checks whether the given object is an instance of the pre-assigned type.
As a result, each object injected by the approach behaves consistently across all \code{isinstance} checks and may pass this check even for project-specific classes.

\paragraph{Super calls.}
Finally, the third change is to replace \code{super} calls with a call to a custom dummy function.
The rationale is that the analyzed function may actually be a method of a class, and the \code{super} call would then refer to the superclass of that class.
However, since we analyze functions in isolation, the surrounding class is not available, and the \code{super} call would always fail.
Instead, our custom dummy function returns the versatile object, and any calls made on the returned object are handled by the standard logic for handling external functions, i.e., \name{} will inject a realistic return value for them.

\subsubsection{Handling Indexing Operations}
\label{sec:indexing ops}

The existing learning-guided execution approach~\cite{fse2023-LExecutor} intercepts variable reads, attribute reads, and function calls to inject predicted values.
In contrast, the existing approach does not handle indexing operations, e.g., \code{x[0]} or \code{x["key"]}.
As a result, indexing operations on values predicted by the neural model often fail, causing the execution to crash.
To avoid such crashes, \name{} extends the existing instrumentation and the neural model to handle indexing operations.
The change in the instrumentation is to intercept any indexing operation, and in case it would usually crash, to query the neural model to predict a value to inject as the result of the indexing operation.
The change in the neural model is to train the model not only on the existing three kinds of values, but also on values that are the result of indexing operations.
As an example, consider a code snippet that reads \code{x["key"]}, where \code{x} was predicted to be a dictionary.
Since the dictionaries injected by the approach are randomly constructed, the \code{"key"} is unlikely to be present in the dictionary, and hence, the code would most likely raise a \code{KeyError}.
To avoid this crash, \name{} queries our updated neural model to predict a value to inject as the result of the indexing operation.

\subsubsection{Covering Different Execution Paths}

The overall goal of \name{} is to find behavioral differences between two versions of a function.
To achieve this goal, it is essential to cover as many different execution paths as possible.
Because the values injected during learning-guided executions are sampled randomly, different executions may follow different paths and trigger different behaviors.
The approach randomizes the injected values in two ways.
At first, the approach picks one of the abstract values in \Cref{tab:values} by sampling from the probability distribution produced by the neural model.
Then, the approach concretizes the selected abstract value as described in \Cref{sec:injecting values}.
We exploit the randomized nature of the injected values by repeating the pairwise learning-guided execution, which probabilistically, causes different values to be injected.
\name{} continues to execute the two functions until it has either identified a behavioral difference, or until reaching a configurable limit of $k=300$ executions.

\subsection{Comparing Execution Behavior}
\label{sec:comparison}

To determine whether the analyzed code change alters the semantics of the function, \name{} compares the execution behavior of the old and the new version of the function.
Intuitively, the approach aims at finding differences in the input-output behavior of the two functions.
More precisely, the approach compares the behavior of the two functions in four ways: by comparing argument and return values, output written to the console, functions that get called, and exceptions.

\subsubsection{Comparing Argument and Return Values}
\label{sec:compare values}

One important aspect in the behavior of a function is the side-effects on the function arguments and on the return values.
\name{} compares these values across the two functions by serializing the value for the old and new function, and by checking whether the serialized values are equal.
If any of the values differ, then the two functions are classified as semantics-changing.
For values that are a versatile object injected by the approach, we adopt an existing approach~\cite{oopsla2015} for recursively flattening an object as a sequence of its attributes, where each attribute is serialized by using its string representation.
For values of other types, the approach uses the string representation of the value.
To allow the comparison of values not only from regular functions but also from generator functions or asynchronous
functions, the approach tries to unwrap the return value before the comparison in case it is a coroutine or generator object.
Additionally, if a return value is a regular functions itself, then the approach tries to execute the function without any arguments, in an effort to compare the behavior of the returned function.

\subsubsection{Comparing Output}

Beyond the return value, the behavior of a function may also be reflected in the output written to the console.
To compare the output of the two functions, \name{} captures the standard output and the standard error of the two functions,
compares the captured outputs, and classifies the two functions as semantics-changing if the outputs differ.

\subsubsection{Comparing Called Functions}
\label{sec:comparing calls}

To accurately identify a change in behavior between the two versions of the function, the approach compares their potential side effects performed by calling external functions.
Because learning-guided executions abstract away the actual behavior of external functions, the approach cannot directly compare these side effects.
Instead, \name{} records a log of all function calls for which the approach injects a return value, along with the arguments passed to these calls.
After both versions of the function have executed, the approach compares their logs of calls to external functions.
If the logs differ, e.g., in the order of called functions or the arguments passed to the functions, the approach classifies the code change as semantics-changing.

\subsubsection{Comparing Exceptional Behavior}
\label{sec:exceptional behavior}

Executing the two functions may cause an exception to be raised in none, one, or both of the functions.
A simple approach could consider the two functions to be semantically equivalent if either both raise an exception or none raises an exception.
However, such an approach would reduce \name{}'s ability to reason about subtle differences in the exception-raising behavior of the two functions.
Specifically, exceptions can be raised for two kinds of reasons.
On the one hand, raising an exception may be part of the intended behavior of the executed code, e.g., to signal an error condition.
Such exceptions are part of the behavior that our approach should compare to determine whether the code change preserves the way in which exceptions are raised.
On the other hand, some exceptions may be caused by the learning-guided execution itself, e.g., when \name{} injects an unrealistic value that the code is not supposed to handle.
Because unrealistic values violate the assumptions of the code, the code may expose arbitrary, possibly exceptional behavior, which should not be compared across the two functions.

To distinguish between exceptions intended by the developer and exceptions caused by our approach itself, we consider two kinds of exceptions to be \emph{intended}.
First, \name{} wraps all exceptions created by a \code{raise} statement with a special class \code{IntentionalException} and modifies any \code{except} clauses around the raise statement to catch \code{IntentionalException}.
Second, the approach considers \code{AssertionError} exceptions as intended exceptions because developers often use them to signal an error condition.

In case at least one of the two functions raises an exception, \name{} compares the raised exceptions by considering the following three scenarios.
(1) One function raises an intended exception, whereas the other does not raise any exception.
In this case, both versions of the function exhibit different semantics and \name{} classifies the code change as semantics-changing.
For example, such a semantic difference may result from a code change that introduces a new error condition or changes the way in which an existing error condition is signaled.
(2) Both functions raise an intentional exception.
In this case, we compare the type of the raised exception and the arguments passed to the exception.
If the exceptions differ, the code change is classified as semantics-changing.
(3) At least one unintended exception is raised, which likely is caused by the approach itself.
In this case, we refrain from drawing any conclusions about this execution.

\subsubsection{Classification of Code Changes}

If any of the above comparison steps reveals a difference in the behavior of the two functions, the approach reports the code change to be \emph{semantics-changing}.
Otherwise, \name{} continues to analyze the code change until exceeding the budget of $k$ repetitions.
If, after exceeding this budget, the approach has not found any behavioral differences, it classifies the code change as follows.
If none of the executions have reached any of the changed code lines, or if all executions finished prematurely due to an unintended exception, then the approach classifies the code change as \emph{inconclusive}.
Otherwise, i.e., \name{} has successfully exercised the changed code but has not found any behavioral differences, the approach classifies the code change as \emph{likely semantics-preserving}.
The reason for saying ``likely'' is that the approach cannot guarantee that the two functions are semantically equivalent, e.g., because the approach may have missed some execution paths.

\section{Implementation}

\name{} builds upon, and significantly extends, the open-source tool LExecutor~\cite{fse2023-LExecutor}.\footnote{\url{https://github.com/michaelpradel/LExecutor/}}
To instrument code, we use the libCST library.\footnote{\url{https://github.com/Instagram/LibCST}}
For training and querying the neural model, we build upon the Transformers library.\footnote{\url{https://github.com/huggingface/transformers}}
The code changes to analyze are automatically extracted from git commits: Given two commit hashes, the approach extracts individual functions that differ between the two commits, and writes the old and new version into a JSON file.
Instead of extracting the changes from commits, one could easily create such a JSON file from other code changes, e.g., changes that are not yet committed to any repository.
The output of the approach is the verdict, i.e., ``semantics-changing'', ``likely semantics-preserving'', or ``inconclusive'', along with a detailed trace of any observed differences in behavior.
For example, if the two executions return different values for the same inputs, the approach prints the inputs and the differing return values.

\section{Evaluation}

Our evaluation addresses the following research questions:
\begin{itemize}
  \item RQ1: How effective is \name{} at classifying code changes?
  \item RQ2: How does \name{} compare to regression testing?
  \item RQ3: How accurate is the neural model underlying the approach?
  \item RQ4: How effective is \name{} at successfully executing the changed code?
  \item RQ5: How efficient is \name{}?
\end{itemize}

\subsection{Experimental Setup}\label{s:setup}

\subsubsection{Dataset for Fine-Tuning}
\label{sec:setup fine-tuning}

Following prior work~\cite{fse2023-LExecutor}, the neural model underlying \name{} is a fine-tuned CodeT5 model~\cite{codet5}.
Because \name{} supports a wider range of values to inject (\Cref{sec:indexing ops}) than prior work~\cite{fse2023-LExecutor}, we fine-tune our own model.
To gather a dataset for fine-tuning, we build upon DyPyBench~\cite{fse2024-DyPyBench}, a benchmark of 50 executable, open-source Python projects.
We use 48 out of the 50 projects, excluding Flask-API and Black to avoid overlap with the projects we evaluate on (\Cref{sec:dataset code changes}).
Using DyPyBench, we manage to collect 250,046 training samples, including 7,502 for the newly added indexing operations (\Cref{sec:indexing ops}).
We shuffle and split the collected data, using 95\% for fine-tuning and 5\% to answer RQ2.
We fine-tune the model for ten epochs.

\subsubsection{Datasets of Code Changes to Analyze}
\label{sec:dataset code changes}

We apply \name{} to four datasets of code changes, which are extracted from ten popular open-source Python projects, listed in \Cref{tab:numberCodeChanges}.

\begin{table}[t]
  \centering
  \caption{Manually annotated code changes.}
  \label{tab:numberCodeChanges}
  \setlength{\tabcolsep}{5pt}
  \begin{tabular}{@{}lrr|rrr@{}}
    \toprule
    Project       & \multicolumn{5}{@{}c@{}}{Code changes}                                                                                       \\
    \cmidrule{2-6}
                  & \multicolumn{2}{@{}c@{}}{By commit message} & \multicolumn{3}{@{}c@{}}{Manually annotated}                                   \\
    \cmidrule{2-6}
                  & \multicolumn{1}{@{}c@{}}{Sem.-preserving}        & Sem.-changing                                     & Sem.-preserving & Sem.-changing & Unclear \\
    \hline
    Airflow       & 31                                          & 15                                           & 15         & 25       & 6       \\
    Black         & 3                                           & 15                                           & 4          & 13       & 1       \\
    FastAPI       & 9                                           & 15                                           & 8          & 12       & 4       \\
    Flask         & 3                                           & 15                                           & 5          & 10       & 3       \\
    HTTPie        & 6                                           & 15                                           & 7          & 14       & 0       \\
    Pandas        & 27                                          & 15                                           & 5          & 13       & 24      \\
    Poetry        & 3                                           & 15                                           & 2          & 11       & 5       \\
    Scikit-Learn  & 37                                          & 15                                           & 23         & 16       & 13      \\
    Scrapy        & 20                                          & 15                                           & 15         & 12       & 8       \\
    TheAlgorithms & 10                                          & 15                                           & 9          & 5        & 11      \\
    \midrule
    Total         & 149                                         & 150                                          & 93         & 131      & 75      \\
    \bottomrule
  \end{tabular}
\end{table}

\paragraph{Manually annotated commits}
To validate the approach against a ground truth, we manually inspect and annotate 299 code changes, as summarized in \Cref{tab:numberCodeChanges}.
As a first step, we filter all commits made before November 1, 2023, based on the following criteria:
(i) the commit is not a merge commit;
(ii) the change affects a single function;
(iii) the commit modifies the function, as opposed to adding or removing an entire function;
(iv) the changed file is a Python file and does not contain ``test'' in its name; and
(v) the old and the new function parse into different ASTs, even after removing comments, decorators, and type annotations.
In an attempt to identify code changes meant to be semantics-preserving, we then filter for commit messages that contain the keyword ``refactor'', ``simplify'', ``cleanup'', or ``optimize'', which yields a total of 149 code changes.
To balance the dataset, we also collect 150 (15 per project) code changes that do not contain the above keywords, and are likely semantics-changing.
Because commit messages alone are not a reliable way of identifying semantics-preserving changes~\cite{murphy}, we then manually inspect each code change and annotate it as either ``semantics-preserving'', ``semantics-changing'', or ``unclear''.
This process results in a ground truth of 93 and 131 code changes annotated as semantics-preserving and semantics-changing, respectively.

\paragraph{Rule-based refactorings}
In addition to developer-created code changes, we also apply \name{} to validate changes made by automated tools.
To this end, we use RIdiom~\cite{ridiom}, a refactoring tool that transforms Python code into a more idiomatic version.
We applying RIdiom to the newer version of the 299 functions in the above dataset and use all of the resulting 165 code changes as an additional dataset.
As RIdiom is designed to make the code more idiomatic, while preserving its behavior, we expect these code changes to be semantics-preserving.

\paragraph{Refactorings created by GPT-3.5 and GPT-4}
Finally, we apply \name{} to code changes created by large language models (LLMs).
LLMs have the potential to support developers by refactoring their code, and we evaluate to what extent \name{} could be used to validate the correctness of such LLM-generated code changes.
Similar to the above setup, we ask two of OpenAI's recent models (gpt-3.5-turbo-0125 and gpt-4-turbo-2024-04-09) to refactor the newer version of the functions from \Cref{tab:numberCodeChanges}.
We use the following prompt:
``You are a Python expert. Improve the quality of this Python code while preserving its behavior and without renaming variables, adding comments, adding a docstring, or adding imports: <code>''
We discard any code changes that do not modify the AST and code changes that split the given function into multiple functions.
This process results in 187 code changes by the GPT-3.5 and 258 code changes by GPT-4.

\subsubsection{Hardware}

All experiments are performed on a machine with an Intel Xeon Silver 4214 CPU (2.0GHz, 12 cores) running Ubuntu 20.04. We use an NVIDIA Tesla P100 (16GB GPU) and a NVIDIA Tesla T4 (16GB GPU) for fine-tuning and inference, respectively.

\subsection{RQ1: Effectiveness}\label{s:rq1}

The following evaluates the effectiveness of our approach at identifying code changes to be semantics-changing or semantics-preserving.
We first present the results on the manually annotated code changes, then discuss the results on the code changes created by RIdiom and LLMs.

\subsubsection{Results on Manually Annotated Code Changes}

\begin{table}[t]
  \centering
  \caption{Effectiveness of \name{} and regression testing on manually annotated ground truth.}
  \setlength{\tabcolsep}{3pt}
  \label{tab:results}
  \begin{tabular}{lr|rrr|rrr}
    \toprule
    \multicolumn{2}{c|}{Ground truth} & \multicolumn{3}{c|}{\name{}} & \multicolumn{3}{c}{Regression testing} \\
    \cmidrule{1-2} \cmidrule{3-5} \cmidrule{6-8}
    & Total & Changing & Preserving & Inconclusive & Changing & Preserving & Inconclusive \\
    \midrule
    Changing & 131 & 91 & 12 & 28 & 10 & 29 & 92 \\
    Preserving & 93 & 27 & 48 & 18 & 2 & 18 & 73 \\
    \midrule
    Total & 224 & 118 & 60 & 46 & 12 & 47 & 165 \\
    \bottomrule
  \end{tabular}
\end{table}

\Cref{tab:results} (center) shows the results of applying \name{} to the 224 code changes that are manually annotated as either semantics-preserving or semantics-changing.
For the task of identifying semantics-changing code changes, the precision is $91/(91+27)=77.1\%$ and the recall is $91/(91+12+28)=69.5\%$.
The overall accuracy across all cases where the approaches gives an answer, i.e., excluding cases where \name{} responds with ``inconclusive'' is $(91+48)/(118+60)=78.1\%$.
That is, \name{}'s predictions are correct for the vast majority of code changes, offering both high precision and high recall.

\begin{figure}
    \begin{lstlisting}
        else: raise AssertionError(
        (@\HLred@)-       "Unexpected IPython magic {node.value.func.attr!r} found. "(@\HLredoff@)
        (@\HLgreen@)+       f"Unexpected IPython magic {node.value.func.attr!r} found. "(@\HLgreenoff@)
                "Please report a bug on https://github.com/psf/black/issues.") from None
    \end{lstlisting}
  \caption{Code change intended to change the semantics, which \name{} confirms.}
  \label{fig:truepositive}
\end{figure}

\Cref{fig:truepositive} shows an example where the developers intend to change the semantics, which \name{} correctly confirms to be true.
The exception raised by the old code had an incorrect message.
The new code fixes this problem by turning the string into an f-string.
Because our approach reaches the exception-triggering code and compares the messages, it confirms that the code change is indeed semantics-changing.\footnote{\url{https://github.com/psf/black/commit/72a84d4099f2930979bd1ca1d9e441140b0a304d}}
Our motivating example from \Cref{fig:motivating} is another code change that \name{} classifies as semantics-changing.
However, in this case the developers did not intend any change in behavior.
This and other examples like it demonstrate that our approach can be useful for detecting unintended changes in behavior.

\begin{figure}
  \centering
  \includegraphics[width=\linewidth]{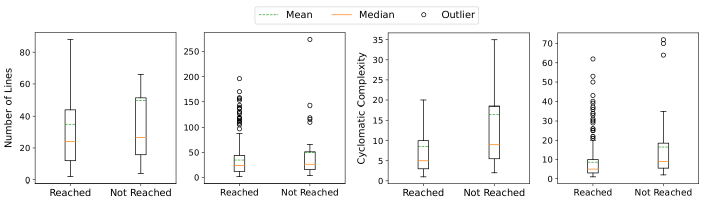}
  \caption{Impact of function size and complexity on \name{}'s ability to reach the changed code lines.}
  \label{fig:boxplot}
\end{figure}

While the majority of code changes is classified correctly, \name{} also has 27 false positives, 12 false negatives, and 46 code changes classified as inconclusive.
We manually inspect these cases to better understand the limitations of our approach.
The main reason for false positives, i.e., code changes classified as semantics-changing even though the change preserves the behavior, is external function calls.
For 21 out of 27 false positives, the code change moves, removes, or changes a function call.
Since our approach reasons about the changed code in isolation, it does not know what side effects, if any, external functions have, and instead classifies any difference in the observed calls as a behavioral difference (\Cref{sec:comparing calls}).
Each of the remaining six false positives has a unique reason, e.g., related to how \name{} compares the argument and return values of the two functions (Section~\ref{sec:compare values}).
False negatives occur when \name{} executes at least parts of the the changed code, but nevertheless cannot observe any behavioral difference.
The main reasons for such cases is the behavioral difference manifests only under specific values, or combination of values, but those values are not predicted by the neural model.
One way to address this limitation could be to combine learning-guided execution with more systematic, constraint-based reasoning about execution paths.
Finally, we also inspect those 46 cases where the approach refrains from making a prediction, but instead responds with ``inconclusive''.
The main reason for such cases is that the approach cannot reach those lines in a large and complex functions that were changed.
To corroborate this hypothesis, \Cref{fig:boxplot} shows how the number of non-comment, non-empty lines of code in a function and the cyclomatic complexity of the function impacts the chance of reaching the modified code lines.
As illustrated by the figure, code changes that are not reached tend to be in larger and more complex functions.

\subsubsection{Results on Automatically Refactored Code}

\begin{table}
  \begin{minipage}[t]{.67\textwidth}
    \centering
    \caption{Effectiveness on automatically refactored code.}
    \label{tab:automatic}
    \begin{tabular}{@{}lrrr@{}}
      \toprule
      Refactoring tool & \multicolumn{3}{c}{Prediction}                         \\
      \cmidrule{2-4}
                        & Inconclusive                   & Changing & Preserving \\
      \midrule
      RIdiom           & 38                             & 5        & 122        \\
      GPT-3.5          & 31                             & 87       & 69         \\
      GPT-4            & 38                             & 143      & 77         \\
      \bottomrule
    \end{tabular}
  \end{minipage}%
  \begin{minipage}[t]{0.33\textwidth}
    \centering
    \caption{Neural model accuracy.}
    \label{tab:accuracyModel}
      \begin{tabular}{@{}lr@{}}
        \toprule
         & Accuracy \\
        \midrule
        Top-1 & 95.13\%  \\
        Top-3 & 96.17\%  \\
        Top-5 & 97.52\%  \\
        \bottomrule
      \end{tabular}
  \end{minipage}
\end{table}

\name{} cannot only reason about manually created code changes, but also about code changes created by automated refactoring tools.
The following reports the effectiveness of the approach on refactorings created via an existing, rule-based approach and via LLMs (\Cref{sec:dataset code changes}).
\Cref{tab:automatic} summarizes the results.

\begin{figure}
    \centering
    \begin{subfigure}[b]{\linewidth}
      \begin{lstlisting}
  (@\HLred@)- provider = Provider(self._pkg, self._pool, self._io)(@\HLredoff@)
  (@\HLred@)- locked = {}(@\HLredoff@)
  (@\HLred@)- for package in self._locked.packages: locked[package.name] = package(@\HLredoff@)
  (@\HLgreen@)+ provider, locked = Provider(self._pkg, self._pool, self._io), {}(@\HLgreenoff@)
  
      \end{lstlisting}
      \caption{Code change performed by RIdiom.}
      \label{fig:ridiomtp}
    \end{subfigure}
    \begin{subfigure}[b]{\linewidth}
      \begin{lstlisting}
  (@\HLred@)- if isinstance(arr_or_dtype, ExtensionType): return arr_or_dtype.name == "category"(@\HLred@)
  (@\HLred@)- if arr_or_dtype is None: return False(@\HLred@)
  (@\HLred@)- return CategoricalDtype.is_dtype(arr_or_dtype)(@\HLred@)
  (@\HLgreen@)+ if isinstance(arr_or_dtype, ExtensionType) and arr_or_dtype.name == "category": return True(@\HLgreenoff@) 
  (@\HLgreen@)+ elif arr_or_dtype is None: return False(@\HLgreenoff@)
  (@\HLgreen@)+ else: return CategoricalDtype.is_dtype(arr_or_dtype)(@\HLgreenoff@)
      \end{lstlisting}
      \caption{Code change performed by GPT-3.5.}
      \label{fig:incorrect_if}
    \end{subfigure}
    \begin{subfigure}[b]{\linewidth}
      \begin{lstlisting}
    def param_allowed(stat_name, include, exclude):
      if not include and not exclude: return True
(@\HLred@)-     for p in exclude:(@\HLredoff@)
(@\HLred@)-         if p in stat_name: return False(@\HLredoff@)
(@\HLred@)-     if exclude and not include: return True(@\HLredoff@)
(@\HLred@)-     for p in include:(@\HLredoff@)
(@\HLred@)-         if p in stat_name: return True(@\HLredoff@)
(@\HLred@)-     return False(@\HLredoff@)  
(@\HLgreen@)+     if any(p in stat_name for p in exclude): return False(@\HLgreenoff@)
(@\HLgreen@)+     if include: return any(p in stat_name for p in include)(@\HLgreenoff@)
(@\HLgreen@)+     return not exclude(@\HLgreenoff@)
    \end{lstlisting}
    \caption{Code change performed by GPT-4.}
    \label{fig:gpt4tp}
  \end{subfigure}
  \caption{Code changes performed by automated refactoring tools, which \name{} finds to unexpectedly change the semantics.}
  \label{fig:ridiomexamples}
\end{figure}

When applying \name{} to the 165 code changes created by RIdiom\cite{ridiom}, the approach confirms 122 code changes to be semantics-preserving, marks 38 code changes as inconclusive, and reports a behavioral difference for five code changes.
As the transformations performed by RIdiom are meant to be semantics-preserving, this results mostly aligns with our expectation.
One code change classified by \name{} as semantics-changing is shown in \Cref{fig:ridiomtp}.
In this example, RIdiom accidentally removes lines during the transformation, presumably due to a bug in the tool's implementation.

The last two rows of \Cref{tab:automatic} show the results of applying \name{} to code changes created by LLMs instructed to refactoring the code.
Perhaps surprisingly, many of the code changes created by the LLMs are classified by \name{} as semantics-changing.
For the GPT-3.5-generated code changes, \name{} identifies 87 out of 187 as semantics-changing and only 69 code changes as semantics-preserving.
Manually inspecting 30 randomly sampled examples out of the 87 code changes flagged as semantics-changing shows that 21 are true positives, i.e., the LLM indeed changes the behavior of the code.
\Cref{fig:incorrect_if} shows an example, where the behavior differs if the \code{isinstance} check passes but the \code{arr\_or\_dtype.name} attribute does not equal \code{"category"}.
For the GPT-4-generated code changes, \name{} classifies 143 code changes, i.e., 54.4\% of all code changes, as semantics-changing.
We again inspected a random sample of 30 of these code changes and find 16 of them to indeed modify the behavior, despite the LLM being instructed to preserve the behavior.
In the example in \Cref{fig:gpt4tp}, GPT-4 tries to simplify the logic of the function.
However, if an empty list is passed as an argument to the \code{state\_name} parameter and the \code{include} parameter and a non-empty list is passed as an argument to the \code{exclude} parameter, the behavior changes.
The remaining 14 code changes are false positives, mostly caused by code changes that modify if and how the code invokes external functions, which \name{} cannot reason about.
Overall, our results suggest that LLMs often fail to improve the code while preserving its semantics, and that \name{} provides an effective means to identify such semantics-breaking code changes.

\subsection{RQ2: Comparison with Regression Testing}
\label{sec:rq regression}

As regression testing currently is the most widely used approach to validate code changes, we compare \name{}'s effectiveness against existing regression test suites.
Such test suites exist for all analyzed projects (\Cref{tab:numberCodeChanges}), except for TheAlgorithms.
We perform the comparison for all 224 code changes that are manually annotated as either semantics-preserving or semantics-changing, i.e., where we have a ground truth to use a reference.
To check whether the existing regressions tests correctly identify a code change as semantics-preserving or semantics-changing, we take a two-pronged approach:
First, we check whether the corresponding commit has any associated continuous integration logs on the GitHub Workflows platform.
If such logs exists, we compare the test execution results for the commits of $f_{old}$ and $f_{new}$.
Second, if we cannot find any continuous integration logs, then we try to execute the tests locally as follows:
1) clone the project repository;
2) check out to the commit under analysis;
3) follow the project's instructions on installing dependencies, building, and running the tests.
Finally, we compare the test execution results for the commits of $f_{old}$ and $f_{new}$.
Because all code changes affect a single, non-test function, the test cases executed for $f_{old}$ and $f_{new}$ are always the same.
Hence, we can directly compare the number of passing and failing tests before and after the code change.
If these numbers are the same, the regression tests consider the code change to be ``semantics-preserving'', and ``semantics-changing'' otherwise.
For code changes without any test execution results, e.g., because the project fails to build, we consider regression testing to be ``inconclusive''.

\Cref{tab:results} (right) shows the results of regression testing on the manually annotated ground truth.
For those code changes that have regression testing results, the tests correctly identify semantics-changing code changes with a precision of $10/(10 + 2) = 83.3\%$, a recall of $10/(10 + 29 + 92) = 7.6\%$, and an accuracy of $(10 + 18)/(12 + 47) = 47.5\%$.
Compared to \name{}, regression testing provides a slightly higher precision ($83.3\%$ vs.\ $77.1\%$), which comes at the expense of a huge drop in recall though ($7.6\%$ vs. $69.5\%$).
The reasons for this low recall include that the existing tests do not cover the changed code, the specific commit cannot be built correctly, or that there simply are no regression tests.
Out of the ten cases that regression testing correctly identifies as semantics-changing, \name{} also identifies seven of them.
Overall, these results show that \name{} is more effective than the existing regression tests at identifying semantics-changing and semantics-preserving code changes, adding benefit over the current state of the art in validating code changes.

\subsection{RQ3: Accuracy of the Neural Model}
\label{sec:rq model}

The following two research questions study two important factors that contribute to the effectiveness of the overall approach, starting with the accuracy of the neural model underlying \name{}.
We evaluate the model, which is trained to predict an abstract value (\Cref{tab:values}) for a given code context, on a held-out subset of all available data (\Cref{sec:setup fine-tuning}).
We report top-$k$ accuracy for different values of $k$, where a prediction is counted as correct if and only if the correct abstract value is among the top-$k$ values predicted by a the model.

Table~\ref{tab:accuracyModel} shows the results.
When considering only the top-most prediction, the model predicts the correct abstract value for 95.1\% of the cases.
The accuracy increases further when considering more predictions, with the top-5 accuracy reaching 97.5\%.
As a point of reference, the top-1 accuracy of prior work~\cite{fse2023-LExecutor} has been 88.1\% in their ``coarse-grained'' prediction task, which is based on the same twelve abstract values as in our work.
There are two main differences.
First, \name{} expands the prediction task to also predict values for indexing operations (\Cref{sec:indexing ops}).
Second, we use a larger and more diverse training dataset, made available by DyPyBench~\cite{fse2024-DyPyBench}.
Because of these differences, a direct comparison of accuracy values is not meaningful.
However, given that the accuracy of our model is high and higher than in the existing learning-guided execution work, we conclude that the neural model is overall successful at predicting realistic values.

\subsection{RQ4: Robustness and Coverage}

The following evaluates \name{}'s ability to execute the changed code, which is a prerequisite for reasoning about the code change.
Specifically, we measure two properties: 
1) We assess the robustness of the approach by measuring how often \name{} \emph{successfully executes} the comparison program, where we count an execution as successful if it does not raise any exceptions, except for intended exceptions and assertion violations.
2) We assess how much of the code in the analyzed functions \name{} executes by measuring \emph{coverage}, which we define as the number of executed lines over all code lines.
As a baseline, we compare against LExecutor, i.e., the state-of-the-art learning-guided execution technique~\cite{fse2023-LExecutor}.
For a comparison with other baselines, e.g., executing code by deterministically or randomly injecting values into the code, or via unit-level test generation, we refer to previous results showing that LExecutor outperforms those baselines~\cite{fse2023-LExecutor}.

\begin{figure}
    \centering
    \begin{subfigure}[]{0.32\linewidth}
      \centering
      \includegraphics[width=\linewidth]{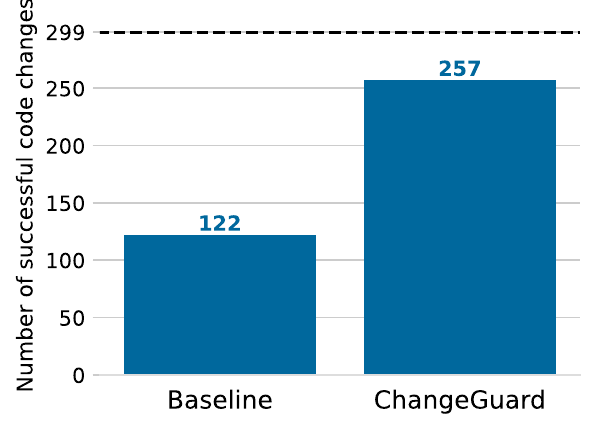}
      \caption{Successfully executed.}
      \label{fig:robustness}
    \end{subfigure}
    \hfill
    \begin{subfigure}[]{0.32\linewidth}
      \centering
      \includegraphics[width=\linewidth]{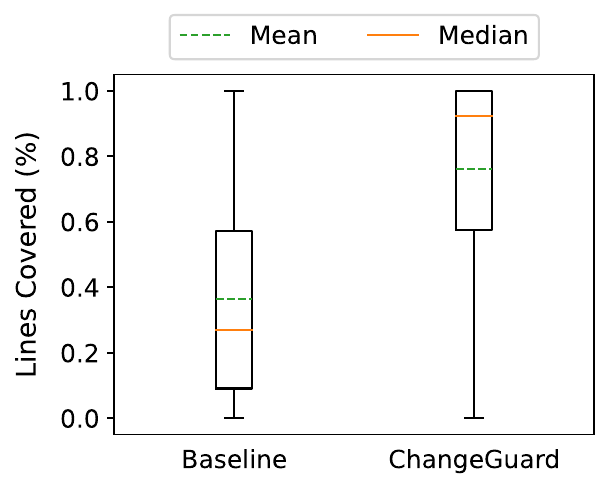}
      \caption{Line coverage.}
      \label{fig:coverageAll}
    \end{subfigure}
    \hfill
    \begin{subfigure}[]{0.32\linewidth}
      \centering
      \includegraphics[width=\linewidth]{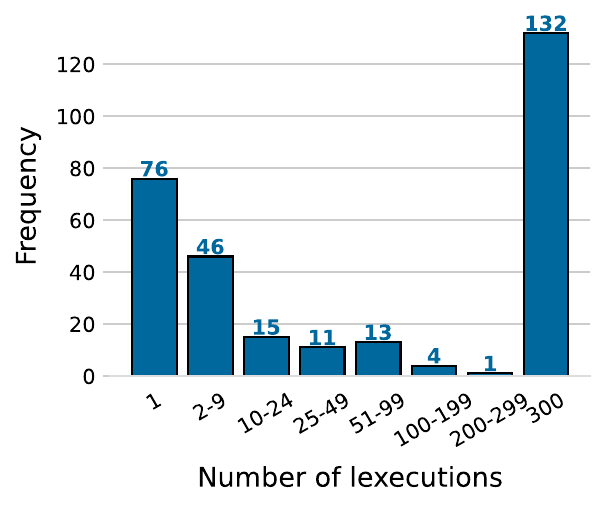}
      \caption{Reaches decision.}
      \label{fig:nb_iterations}
    \end{subfigure}

    \caption{Robustness and coverage compared to the baseline learning-guided execution~\cite{fse2023-LExecutor}, and number of executions before \name{} reaches a decision, across all 299 manually annotated code changes.}
    \label{fig:robustnessCoverage}
    \vspace{-3mm}
\end{figure}

\Cref{fig:robustnessCoverage} summarizes the results.
As shown in \Cref{fig:robustness}, \name{} successfully executes 257 out of 299 code changes, enabling it to meaningfully compare the large majority of all code changes.
This result significantly improves upon the baseline, which can successfully execute only 122 out of the 299 code changes.
The fact that \name{} more than doubles the robustness of the approach compared to the baseline shows the importance of the improvements described in \Cref{sec:pairwise lexecution}.
As shown in \Cref{fig:coverageAll}, these improvements also increase the coverage achieved by the approach.
With a median line coverage of 92\%, \name{} exercise the large majority of all code in the analyzed functions.
Overall, the high robustness and coverage of our approach is an important ingredient for \name{}'s effectiveness.

\subsection{RQ5: Efficiency}

To quantify how efficient \name{} is at determining whether a code change is semantics-preserving or semantics-changing, we measure the time it takes to instrument and execute the comparison program.
Instrumenting the comparison program takes 1.15 seconds, on average.
Once instrumented, \name{} repeatedly executes the comparison program until finding a behavioral difference.
Because our implementation caches requests to the neural model, most requests to the model happen during the first execution, and we hence measure the time taken by the first and later executions separately.
On average, the first execution of the comparison program takes 1.67 seconds, and the remaining executions each take 1.01 seconds.

The overall time for classifying a code change depends on how often the approach repeats the pairwise learning-guided execution.
\Cref{fig:nb_iterations} shows that the number of executions required to reach a conclusion follows a bimodal distribution: either the approach quickly identifies that the old and the new function have diverging behavior (left end of the plot), or the approach keeps executing the comparison program without finding any difference (right end of the plot).
More precisely, for 76 out of the 299 code changes the approach only needs one execution to identify a change in semantics.
Increasing the number $k$ of repetitions yields diminishing returns, with only a single behavioral difference found in executions 200 to 299.
In other words, by reducing the number $k$ of executions, one can trade efficiency for a small reduction in recall.

\section{Limitations, Threats to Validity, and Future Work}

Our approach has several limitations that affect the generalizability of our results.
First of all, our implementation targets Python, which, as a highly dynamic language, is particularly attractive for a dynamic analysis approach, but also facilitates the implementation of learning-guided execution.
Learning-guided execution has so far been implemented for Python only~\cite{fse2023-LExecutor,icse2025-Treefix}. Applying it to another highly dynamic language, e.g., JavaScript, seems relatively straightforward, whereas extending learning-guided execution, and by extension also \name{}, to statically typed languages will pose interesting new research challenges.

Second, our work focuses on single-function changes.
Reasoning about changes that span multiple functions will require defining ``semantics-preserving'' for more complex code and new techniques, e.g., to select an entry point for the execution.
Going in the opposite direction, future work could also explore to adopt \name{} to code snippets smaller than a single function.
For example, if a single expression or a single statement is changed, \name{} could be applied only to the changed code area to check whether its semantics has changed.

As a third limitation, our approach does not consider the side effects of external functions, which is the main cause of false positives (Section~\ref{s:rq1}).
This limitation could be addressed in the future either by running \name{} in an environment with all dependencies installed or by modeling the side effects of external functions.

Fourth, our approach assumes the changed code to behave deterministically, which has been the case for all code changes in our evaluation. 

Finally, our work inherits the general limitation of learning-guided execution to not guarantee executions to be realistic.
\name{} mitigates this limitation through  several novel techniques that make learning-guided execution more robust (Section~\ref{sec:injecting values}--\ref{sec:indexing ops}) and by training a highly accurate neural model (Section~\ref{sec:rq model}).

\section{Related Work}

\paragraph{Reasoning about code changes}
Motivated by the importance of code changes, various approaches try to predict their risk~\cite{risk, risk2}, e.g., via just-in-time defect prediction~\cite{DBLP:journals/csur/ZhaoDC23,Hoang2019, Yang2015a,cc2vec}.
While ``risk'' and ``defect'' are defined broadly in prior work, we focus on the orthogonal problem of validating whether a code change modifies the runtime behavior of the code.
Another line of work tries to identify breaking API changes, e.g., via static analysis~\cite{brito2018apidiff} or by observing type signatures during testing~\cite{mezzetti2018type}.
\name{} differs by exercising the changed code in a targeted manner and by performing a detailed comparison of the runtime behavior.
DiffSearch~\cite{diffsearch} is a search engine to find specific code changes.
Our work differs from all the above work by using learning-guided execution to reason about code changes.

\paragraph{Checking code for equivalence}
Even though the problem of deciding whether two pieces of code are equivalent is undecidable in general, many approaches try to address it approximately.
Unlike approaches that compare functions at the binary level~\cite{binary, binary2}, our approach works on source code and executes the code to observe its behavior.
Clone detection~\cite{clones, clones2, clones3}, especially approaches that can detect type-4 clones~\cite{type4, type42, type43, type44} also rely on static analysis.
EQMiner mines functionally equivalent code fragments via random testing~\cite{Jiang2009a}.
\name{} differs by using learning-guided execution instead of purely random testing, by reasoning about code changes, and by considering side-effects of external functions when comparing the execution behavior.

\paragraph{Refactorings}
The ability of \name{} to identify semantics-changing code changes can be used to validate the transformations performed by automated refactoring tools.
We explore this usage scenario in our evaluation with code changes created by RIdiom~\cite{ridiom} and by large language models.
Another refactoring-related line of work is the automatic detection of refactorings~\cite{fowler}, e.g., Ref-Finder~\cite{refactoring}, RefDiff~\cite{refactoring2, refactoring22}, RefactoringCrawler~\cite{refactoring3}, and RefactoringMiner~\cite{refactoring4, refactoring42}.
These techniques are all based on static analysis and detect semantics-preserving changes by looking for specific patterns, whereas \name{} dynamically executes the code before and after a potential refactoring to observe its behavior.

\paragraph{Testing and automatic test generation}
The currently most common way to validate code changes in practice is through regression testing.
We show in our evaluation that \name{} significantly increases the ability to find behavior-changing code changes compared to the existing regression tests (Section~\ref{sec:rq regression}).
Learning-guided execution and automated test generation share the goal of executing code.
Several test generators for Python have been proposed recently, e.g., Pynguin~\cite{pynguin}, CodaMosa~\cite{generation2}, CoverUp~\cite{generation3}, and SymPrompt~\cite{generation1}.
Test generators typically assume that the code under test comes with all dependencies available, and they rely on a test oracle to validate the behavior.
Instead, \name{} executes incomplete code in isolation, and it compares the code before and after a change.
We refer to prior work~\cite{fse2023-LExecutor} for an empirical comparison of learning-guided execution with test generation.

\section{Conclusion}

This paper introduces \name{}, a novel approach for identifying semantics-breaking code changes using pairwise learning-guided execution.
Our evaluation on a diverse set of code changes in popular Python software shows high precision and recall in detecting unintended behavioral modifications, and it demonstrates that \name{} enhances the robustness and coverage of the state-of-the-art existing learning-guided execution technique.
We envision \name{} to serve as a validation step both for code changes made by developers and by automated code transformation tools, to ensure that the behavior of the code is preserved.

\section*{Data Availability}

Our code and data are available at
\url{https://github.com/sola-st/ChangeGuard/}.

\section*{Acknowledgments}

This work was supported by the European Research Council (ERC, grant agreements 851895 and 101155832) and by the German Research Foundation within the ConcSys, DeMoCo, and QPTest projects.

\bibliographystyle{ACM-Reference-Format}
\bibliography{referencesMore,referencesMP}

\end{document}